# Strong Band Hybridization between Silicene and Ag(111) Substrate

Yakun Yuan,[1,†] Ruge Quhe,[1,2,†] Jiaxin Zheng,[1,2] Yangyang Wang,[1] Zeyuan Ni,[1] Junjie Shi,[1] and Jing Lu[1,*]

[1]State Key Laboratory of Mesoscopic Physics and Department of Physics, Peking University, Beijing 100871, P. R. China

[2]Academy for Advanced Interdisciplinary Studies, Peking University, Beijing 100871, P. R. China

[†]These authors contributed equally to this work.

*Corresponding author: jinglu@pku.edu.cn

**Abstract**

By using first-principles calculations, we systematically investigated several observed phases of silicene on Ag(111) substrates (($2\times2$)silicene/($\sqrt{7}\times\sqrt{7}$)Ag(111), ($\sqrt{7}\times\sqrt{7}$)silicene/($2\sqrt{3}\times2\sqrt{3}$)Ag(111), ($\sqrt{7}\times\sqrt{7}$)silicene/($\sqrt{13}\times\sqrt{13}$)Ag(111)) and their electronic structures. We find that the original Dirac cone of silicene is about 1.5-1.7 eV deeply below the Fermi level and severely destroyed by the band hybridization between silicene and Ag in all the examined phases. Thus, silicene synthesized on Ag(111) substrates could not preserve its excellent electronic property and new method is needed to develop in synthesizing silicene with its Dirac cone surviving.

**Introduction**

Analogous to graphene, silicene is formed by Si atoms, another element in IV, arranged in a honeycomb structure [1]. According to density functional theory (DFT) calculations, two atoms in the unit cell of free-standing silicene have different height along the $z$ axis, forming a low-buckled structure [2,3]. Si-Si bond is formed by partial $sp^3$ hybridization, which results in $\sigma$ bands, while the hybridization of remaining $p_z$ orbitals of Si atoms result in $\pi(\pi^*)$ bands, which have linear dispersion around the Fermi level ($E_f$) and form so-called Dirac cone structure same as graphene. However, due to the stronger spin-orbit coupling (SOC) of Si atom than C atom as well as more connectivity with modern Si-based device technologies, silicene is more eligible in realizing quantum spin Hall effect (QSHE) and fabricating spintronics devices [4,5]. Besides, unlike graphene (single layer), whose band gap is difficult to open without degrading its electronic property [6], theoretical calculations predicted that the band gap of silicene can be opened and tuned by an external perpendicular electric field [7,8] or single-side adsorption of alkali metal atoms [9] with its electronic property preserved because the equivalency of the two sublattices in a low-buckled structure are destroyed by a perpendicular external or built-in electric field.

Unfortunately, silicene is much less stable than graphene and much more difficult to synthesis. So far, in experiment, silicene is obtained by growing on substrate surfaces, such as Ag [3,10-18], $ZrB_2$ [19], and Ir [20], using molecular beam epitaxy (MBE). On Ag(111) substrates, several phasess of silicene have been synthesized with super-periodicity of ($2\times2$)silicene/($\sqrt{7}\times\sqrt{7}$)Ag(111) [11], ($\sqrt{7}\times\sqrt{7}$)silicene/($2\sqrt{3}\times2\sqrt{3}$)Ag(111) [3,12,17,18], ($\sqrt{7}\times\sqrt{7}$)silicene/($\sqrt{13}\times\sqrt{13}$)Ag(111) [16-18], (3×3)silicene/(4×4)Ag(111) [10,13,16-18], and ($\sqrt{3}\times\sqrt{3}$)silicene [14,15]. In early experiments, the observed linear dispersion in the angular-resolved photoelectron spectroscopy (ARPES) is ascribed to the Dirac cone in the silicene energy bands [10,19]. However, later calculations in (3×3)silicene/(4×4)Ag(111) and ($\sqrt{7}\times\sqrt{7}$)silicene/($\sqrt{13}\times\sqrt{13}$)Ag(111) system suggest that the linear dispersion structure observed is mainly contributed by Ag substrate and the Dirac cone is absent in the composite system [13,21-23]. The scanning tunneling spectroscopy (STS) in (3×3)silicene/(4×4)Ag(111) and ($\sqrt{7} \times \sqrt{7}$)silicene/($\sqrt{13} \times \sqrt{13}$)Ag(111) phases confirmed that the silicene layer on Ag(111) has lost its Dirac fermion character and two dimensional character due to the absence of Landau level under a strong magnetic field.

In this Article, by using DFT method, we systematically calculate the geometric and electronic properties of ($2 \times 2$)silicene/($\sqrt{7} \times \sqrt{7}$)Ag(111), ($\sqrt{7} \times \sqrt{7}$)silicene/($2\sqrt{3} \times 2\sqrt{3}$)Ag(111), and ($\sqrt{7} \times \sqrt{7}$)silicene/($\sqrt{13} \times \sqrt{13}$)Ag(111) phases. The electronic structures of the two phases (($2 \times 2$)silicene/($\sqrt{7} \times \sqrt{7}$)Ag(111) and ($\sqrt{7} \times \sqrt{7}$)silicene/($2\sqrt{3} \times 2\sqrt{3}$)Ag(111)) have not been studied so far to the best of our knowledge. We find a strong hybridization between silicene and Ag substrates, which lead to a severe destruction of the Dirac cone of silicene in these phases. Thus, the Dirac fermion is absent in these phases of silicene on Ag substrates, and it is unlikely to observe the predicted exotic properties of silicene.

**Computational details**

Geometry optimizations are performed with ultrasoft pseudopotential [24] plane-wave basis method, implemented in the CASTEP package [25]. The plane-wave cut-off energy is 330 eV. Generalized gradient approximation (GGA) of Perdew-Burke-Ernzerhof (PBE) [26] is used to describe the exchange-correlation effect. A DFT-D semiempirical dispersion-correction approach is adopted to consider the dispersion interaction between silicene and Ag substrates [27]. All the composite structures are established with accurate lattice constant of bulk Ag, and five layers of Ag atoms are used to represent the Ag substrates. During the optimization, the bottom three layers of Ag atoms, the cell shape, and the lattice constants are fixed. To eliminate the spurious interaction between periodic images in the *z* direction, the vacuum space is set no less than 17 Å, and dipole correction is used as well. Only gamma point is considered in *k*-point mesh in the Brillouin zone. The components of the energy band are analyzed by using Vienna *ab initio* simulation package [28] with plane wave cut-off energy of 375 eV and 0.01 Å$^{-1}$ (Fig. 2b), 0.02 Å$^{-1}$ (Fig. 2c,d) Monkhorst-Pack $k$-point mesh [29] sampled in the Brillouin zone. Ultrasoft pseudopotential is also used to describe the inner electrons of atoms. The exchange-correlation function is chosen as GGA-PBE, and dipole correction is also adopted.

**Results and Discussions**

The optimized structures of silicene on Ag substrates are shown in Fig. 1. Due to the presence of Ag substrates, the geometric configurations of silicene are significantly modified. Silicene in $(2\times2)$silicene/$(\sqrt{7}\times\sqrt{7})$Ag(111) and $(\sqrt{7}\times\sqrt{7})$silicene/$(\sqrt{13}\times\sqrt{13})$Ag(111) phases evolve into a 3-layer structure, as shown in Fig. 1a and Fig. 1c, respectively. In $(\sqrt{7}\times\sqrt{7})$silicene/$(2\sqrt{3}\times2\sqrt{3})$Ag(111) phase (Fig. 1b), only 2 of the 14 Si atoms in a super cell remain in the first layer while the other 12 Si atoms stay in the second layer. The maximum buckling of silicene is 2 ~ 3 times larger than that (0.44 Å [2,3]) of the free-standing situation, reaching 1.16 Å, 1.53Å, 1.65 Å for $(2\times2)$silicene/$(\sqrt{7}\times\sqrt{7})$Ag(111), $(\sqrt{7}\times\sqrt{7})$silicene/$(2\sqrt{3}\times2\sqrt{3})$Ag(111), and $(\sqrt{7}\times\sqrt{7})$silicene/$(\sqrt{13}\times\sqrt{13})$Ag(111) phases, respectively. The vertical distance from the bottom layer of silicene to Ag(111) surface is 1.39 Å (Fig. 1a), 1.87 Å (Fig. 1b), 1.23 Å (Fig. 1c), respectively, indicating a chemical adsorption of silicene on Ag substrates. We also calculate the binding energy $E_b$ of the above three structures, which is defined as below:

$$E_b = (E_{Si} + E_{Ag} - E_{Si/Ag})/N$$

Where $E_{Si}$, $E_{Ag}$ and $E_{Si/Ag}$ are the energy for free-standing silicene, clean Ag(111) substrates, and composite systems, respectively, and $N$ is the number of silicon atoms in one unit cell. We find the binding energy $E_b$ for the three phases is 0.66 ($(2\times2)$silicene/$(\sqrt{7}\times\sqrt{7})$Ag(111)), 0.77 ($(\sqrt{7}\times\sqrt{7})$silicene/$(2\sqrt{3}\times2\sqrt{3})$Ag(111)), and 0.68 eV/Si atom $((\sqrt{7}\times\sqrt{7})$silicene/$(\sqrt{13}\times\sqrt{13})$Ag(111)), which are comparable with the theoretical binding energy value of 0.67 eV/Si atom for $(\sqrt{7}\times\sqrt{7})$silicene/$(2\sqrt{3}\times2\sqrt{3})$Ag(111) and 0.72 eV/Si atom for $(\sqrt{7}\times\sqrt{7})$silicene/$(\sqrt{13}\times\sqrt{13})$Ag(111) phases [30] and manifests that silicene is strongly chemisorbed on Ag(111) substrates.

The band structures for the above phases are shown in Fig. 2. For clarification, we project the electron states onto silicene (red) and Ag substrates (gray). The energy band of free-standing silicene (Fig. 2a) is also plotted for comparison. From the energy bands of $(2\times2)$silicene/$(\sqrt{7}\times\sqrt{7})$Ag(111) phase (Fig. 2b), we observe a structure analog to the Dirac cone about 1.5 eV below $E_f$ around $K$ point, which has characteristics of the $\pi(\pi^*)$ bands in free-standing silicene as shown in Fig. 2a, but apparent Ag component is also visible in this cone. Similarly, there exist silicene contributed bands around $K$ point and about 1.6 and 1.7 eV below $E_f$ in $(\sqrt{7}\times\sqrt{7})$silicene/$(2\sqrt{3}\times2\sqrt{3})$Ag(111) and $(\sqrt{7}\times\sqrt{7})$silicene/$(\sqrt{13}\times\sqrt{13})$Ag(111) phases, respectively, which also probably originate from the $\pi(\pi^*)$ bands of silicene. However, these states nearly lose the Dirac cone shape. Therefore the strong band hybridization between silicene and Ag substrates in the above three phases leads to severe modification or destruction of the Dirac cone of silicene, and Dirac fermions are absent in these phases.

In order to further explore the hybridization in the three phases, we investigate the total electron density as well as the electron density of several critical states in the band structure. Contour plots of total electron density are shown in Fig. 3. An accumulation of electron density between some Si and surface Ag atoms is observed in the three phases. This could serve as an evidence for the formation of covalent bonds between corresponding Si and Ag atoms in addition to the ionic interaction, which leads to the strong hybridization in the energy band of above three phases.

We plot electron density distribution of the critical states which has contribution from the states of silicene's $\pi$ and $\pi^*$ band on the $K$ point in Fig. 4 and Fig. 5, respectively. For comparison, the corresponding states in the $\pi$ and $\pi^*$ band on the $K$ point of free-standing silicene are shown in Fig. 4a and Fig. 5a, respectively. The electron density of both the $\pi$ and $\pi^*$ states in $(2\times2)$silicene/$(\sqrt{7}\times\sqrt{7})$Ag(111) phase is most similar to that of pure silicene (Fig.4b and Fig. 5b), exhibiting a feature in agreement with the relatively better maintenance of the Dirac cone in this phase (Fig. 2b). But the two states are largely extended to Ag substrate, which is consistent with the strong mixing of Si and Ag components in the band structure. By contrast, both the $\pi$ and $\pi^*$ states of ($\sqrt{7}\times\sqrt{7}$)silicene/($2\sqrt{3}\times2\sqrt{3}$)Ag(111) (Fig. 4c and Fig. 5c) and ($\sqrt{7}\times\sqrt{7}$)silicene/($\sqrt{13}\times\sqrt{13}$)Ag(111) phases are largely distorted, which is in agreement with the severe distortion of the Dirac cone in the band structures in these phases (Fig. 2c and 2d). The two states have a large distribution on Si atoms, exhibiting a character consistent of the more Si component than Ag around the Dirac point in the band structures. The distribution of the silicene $\pi$ and $\pi^*$ states around the $K$ point provides another evidence of the existence of strong hybridization between silicene and Ag substrates.

In conclusion, due to the higher chemical reactivity of silicene compared with graphene, there is a severe hybridization between silicene and Ag(111) substrates in ($2\times2$)silicene/($\sqrt{7}\times\sqrt{7}$)Ag(111), ($\sqrt{7}\times\sqrt{7}$)silicene/($2\sqrt{3}\times2\sqrt{3}$)Ag(111), and ($\sqrt{7}\times\sqrt{7}$)silicene/($\sqrt{13}\times\sqrt{13}$)Ag(111) phases. As a result, the Dirac cone structure in the energy band of above three phases is substantially modified or destroyed and Dirac fermions no longer exist in those phases. Thus, current silicene synthesized on Ag(111) substrates could not preserve its excellent electronic property and new method is still needed to be proposed in synthesizing silicene with Dirac fermions.

**Acknowledgement**

This work was supported by the National Natural Science Foundation of China (Nos. 11274016, 51072007, 91021017, 11047018 and 60890193), the National Basic Research Program of China (Nos. 2013CB932604 and 2012CB619304), Fundamental Research Funds for the Central Universities, National Foundation for Fostering Talents of Basic Science (No. J1030310/No. J1103205), Program for New Century Excellent Talents in University of MOE of China. J. Zheng also acknowledges the financial support from the China Scholarship Council. J.L. thanks Prof. K. H. Wu helpful discussions.

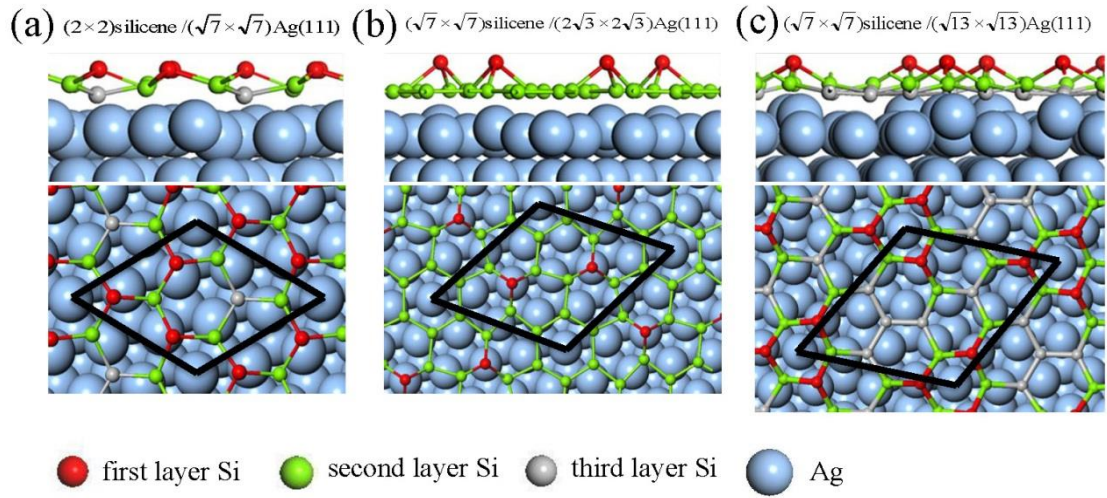

**Figure 1: Side view (top parts) and top view (bottom parts) of the optimized geometry structures of silicene on Ag(111) substrates.** The maximum buckling of silicene in these composite structures are 1.16 Å, 1.53 Å, and 1.65 Å, respectively. The shortest distances between the bottom silicon atoms and Ag(111) surfaces are 1.39 Å (a), 1.87 Å (b), and 1.23 Å (c).

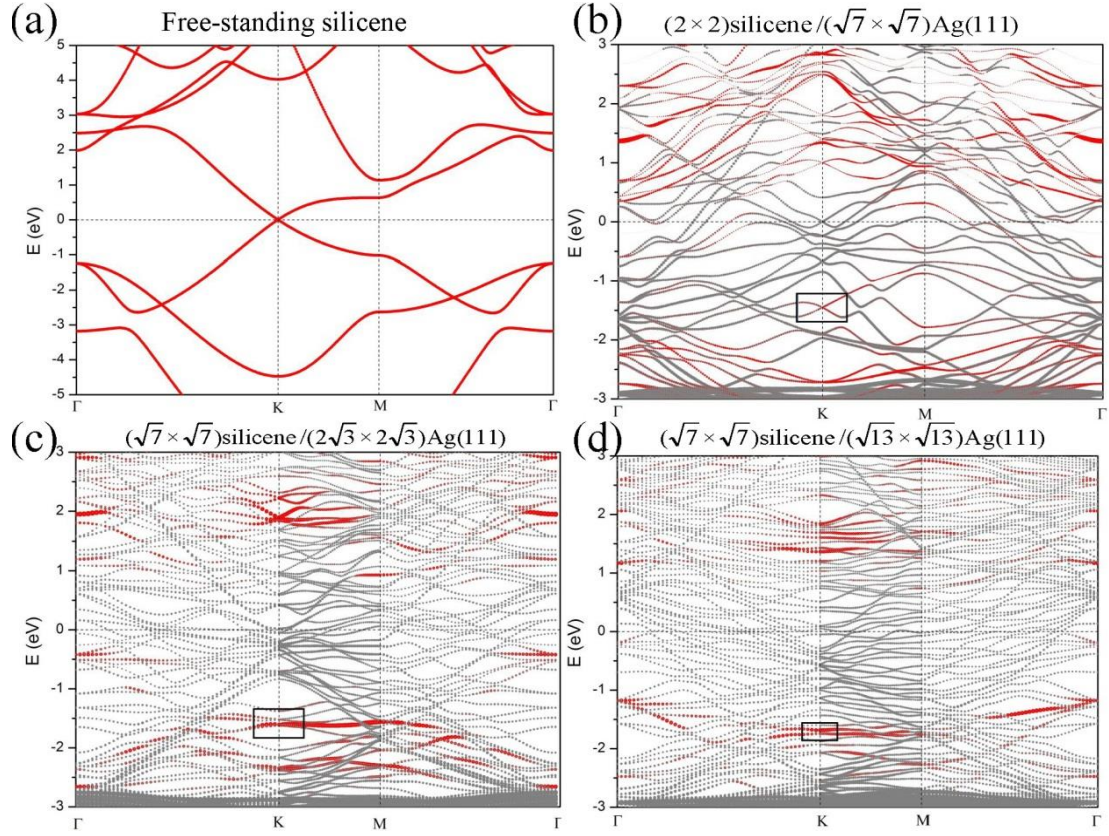

**Figure 2:** (a) **Free-standing silicene band structure, in which the $\pi$ and $\pi^*$ bands degenerate at the Fermi level.** (b-c) **Projected band structures of silicene/Ag(111) composite systems.** Red dots and gray dots represent the states contributed by Ag and Si atoms, respectively. The intensity of the color is proportional to the weight of the corresponding atoms. Black squares indicate those states probably contributed by the $\pi(\pi^*)$ bands of silicene. The Fermi level is set to zero.

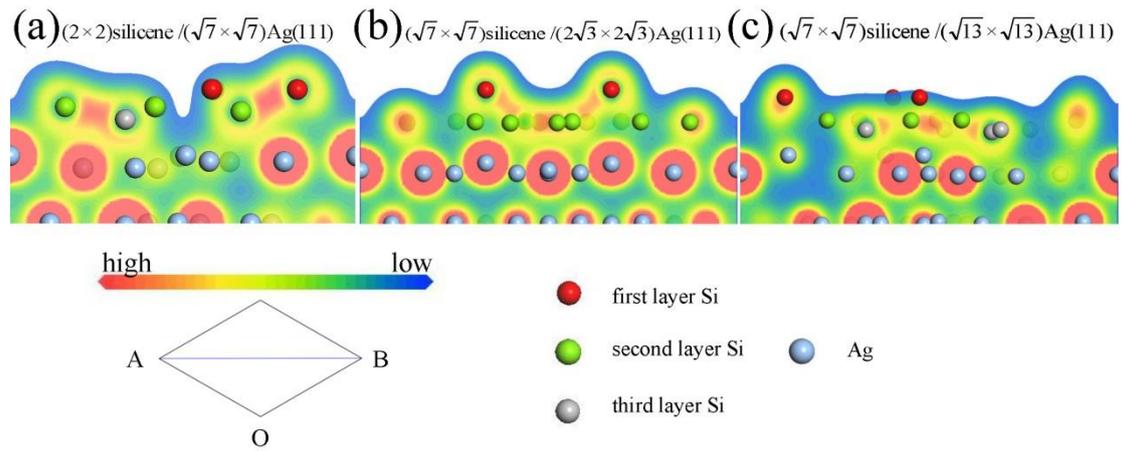

**Figure 3: Contour plots of total electron density in silicene/Ag(111) composite structures.** All the section planes are parallel to the $z$ axis and line AB in super cells as shown in the bottom left plot. The super cells are defined in the bottom panels in Fig. 1 for each structure. Transparent balls indicate the corresponding atoms are behind the section plane.

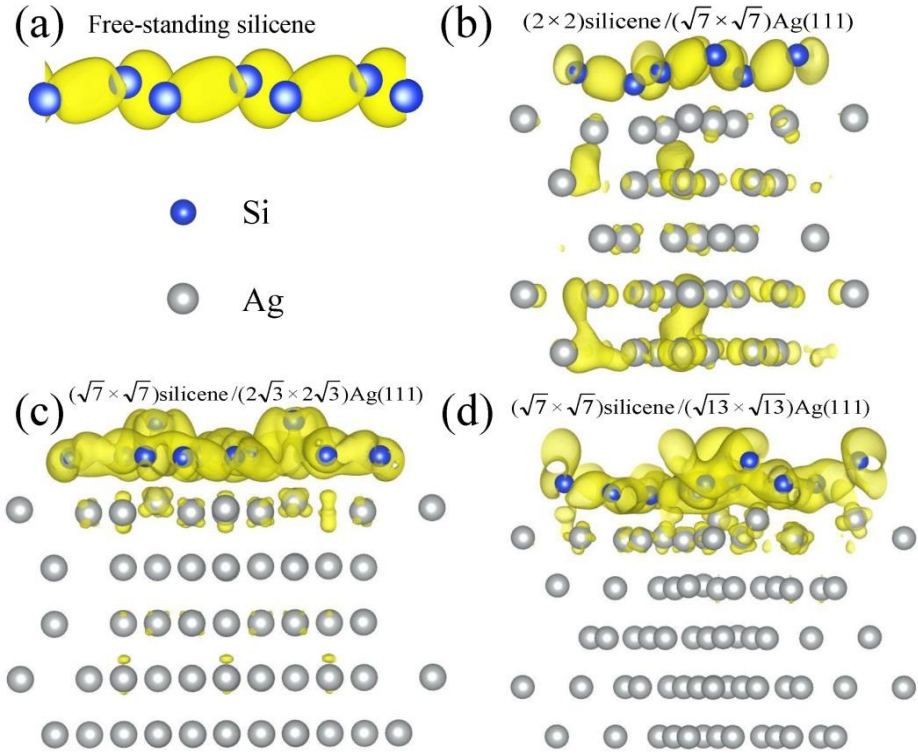

**Figure 4: Isosurface plots of electron density distribution of the critical states which has contribution from the states of silicene's $\pi$ band on the $K$ point.** Due to the existence of Ag substrates, a modification of the $\pi$ characteristics of free-standing silicene states and electron density distributed on Ag substrate are found in the composite structures.

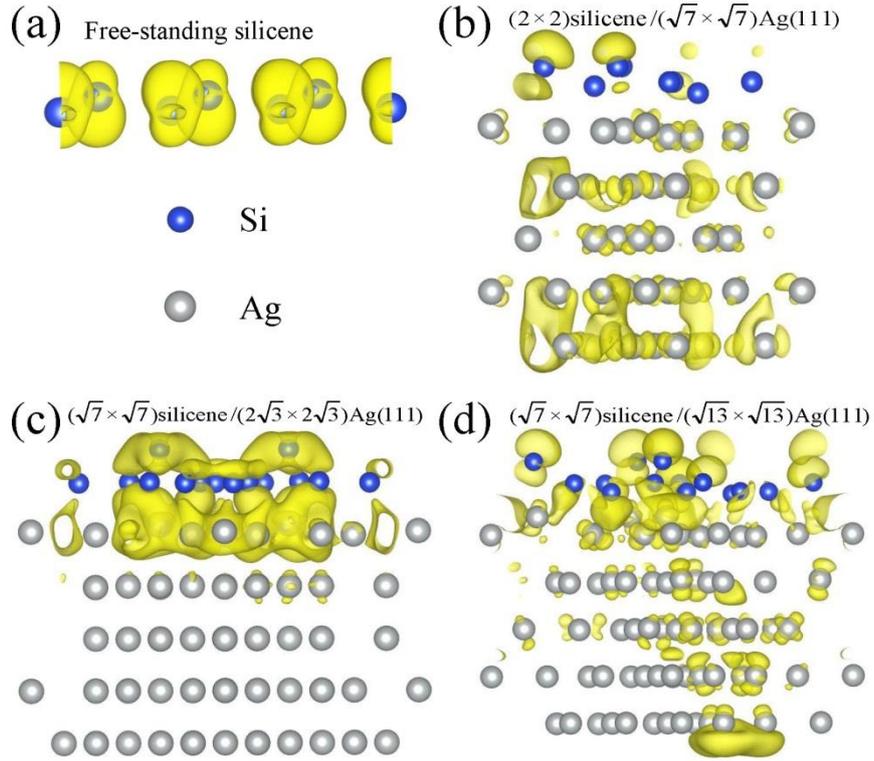

**Figure 5: Isosurface plots of electron density distribution of the critical states which has contribution from the states of silicene's $\pi^*$ band on the *K* point.** Due to the existence of Ag substrates, a modification of the $\pi^*$ characteristics of free-standing silicene states and electron density distributed on Ag substrate are found in the composite structures.